\documentclass[twocolumn,showpacs,preprintnumbers,amsmath,amssymb]{revtex4}

\usepackage{graphicx}
\usepackage{dcolumn}
\usepackage{bm}
\usepackage{textcomp}
\usepackage{wasysym}
\usepackage{stmaryrd}

\def \ee{\end{equation}}
\def \be{\begin{equation}}

\preprint{}

\begin{document}

\title{A geometrical dual to 
relativistic Bohmian mechanics - the
multi particle case}

\keywords      {Quantum Mechanics, Curved space-time}
\author{Benjamin Koch}
 \affiliation{
 Pontificia Universidad Cat\'{o}lica de Chile, \\
Av. Vicu\~{n}a Mackenna 4860, \\
Santiago, Chile \\
}
\date{\today}

\begin{abstract}
In this article it is shown that the
fundamental equations of relativistic Bohmian mechanics
for multiple bosonic particles have a dual description
in terms of a classical theory of curved space-time.
We further generalize the results to interactions 
with an external electromagnetic
field, which corresponds to the minimally coupled Klein-Gordon equation.
\end{abstract}

\pacs{04.62.+v, 03.65.Ta}
\maketitle

%
\section{Introduction}

Understanding quantum mechanics as it is usually taught
is a mayor challenge for many physics students.
For most of them the problem is not 
the introduction of new mathematical tools,
but the understanding of the new concepts like observables.
In standard quantum mechanics observables and the
corresponding uncertainty are promoted to a fundamental
principle, which can not be further understood. 
But it was shown by David Bohm that this does not
necessarily have to be the case \cite{Bohm:1951,Bohm:1951b}.
In the de Broglie-Bohm (dBB) interpretation 
it is explained that the uncertainty ``principle'' and the
description by means of operators can be understood
in terms of uncontrollable initial conditions and non-local
interactions between an additional field (the ``pilot wave'')
and the measuring apparatus. 
This theory was further generalized to relativistic quantum mechanics
and quantum field theory with bosonic and fermionic fields
\cite{Holland:1985ud,Nikolic:2002mi,Struyve:2006cj,Nikolic:2006az}.
It is well known that (due to its non-locality) the dBB theory is not
in contradiction to the Bell inequalities \cite{Bell:1988}.
Due to its contextuality, the dBB theory is also not
affected by the Kochen-Specker theorem \cite{Kochen:1976}.

One mayor drawback of the dBB theory is that the
pilot wave and the corresponding ``quantum potential''
have to be imposed by hand without further justification.
In a previous work \cite{Koch:2008hn,Koch:2008rh} 
it was shown that the 
relativistic dBB theory for a single particle is dual
to a scalar theory of curved spacetime.
In this dual theory the ominous ``pilot wave'' can be readily
interpreted as a well known physical quantity, namely a space-time
dependent conformal factor of the
metric.
This work on the single particle has many features
in common with other publications on the subject
\cite{Santamato:1984qe,Shojai:2000us,Carroll:2004hs,Carroll:2007zh,
Abraham:2008yr}.
However, having a duality for the single particle case
is not enough, because the dBB interpretation only is
a consistent quantum theory when it also has the many particle case. The many
particle theory is for example crucial for understanding the
quantum uncertainty and for evading the non existence theorems
\cite{Bell:1988,Kochen:1976}.
Therefore, we will generalize the previous results and 
present a dual for the relativistic many particle dBB theory. 

The paper is organized a follows:
In the second section a summary of the ingredients
of the many particle dBB theory will be given.
In the third section we will define a theory of curved space-time
and proof that it is dual to the many particle dBB theory.
In the fourth section it is shown, how
both theories can be coupled to an
external electromagnetic field.
The fifth section contains the non-relativistic limit 
(the Schr{\"o}dinger equation). 
It is further discussed what this limit
corresponds to in the dBB theory and
in its geometrical dual.
In the sixth section 
issues like non-locality,
differences to general relativity,
and the universality of the given
matching conditions are addressed.
In the seventh section the results are summarized.
%
\section{Relativistic Bohmian mechanics for many particles}

In this section we shortly list the ingredients for
the interpretation of the many particle quantum Klein-Gordon equation
in terms of Bohmian trajectories. 
For a detailed description of subsequent topics in the dBB 
theory like particle creation,
the theory of quantum measurement,
many particle states, and quantum field theory
the reader is referred to \cite{Nikolic:2006az}.
Let $|0\rangle$ be the vacuum and $|n\rangle$ be an arbitrary n-particle state.
The corresponding n-particle wave function is \cite{Nikolic:2002mi}
\be\label{eq_wavefunction}
\psi(x_1;\;\dots\; ; x_n)= \frac{{\mathcal{P}}_s}{\sqrt{n!}}
\langle 0|\hat{\Phi}(x_1) \dots \hat{\Phi}(x_n)|n\rangle\quad,
\ee
where the $\hat{\Phi}(x_j)$ are scalar Klein-Gordon field operators 
and the symbol ${\mathcal{P}}_s$ denotes symmetrization over all 
positions $x_j$.
For free fields, the wave function (\ref{eq_wavefunction})
satisfies the equation
\be\label{eq_KG0}
\left(\sum_j^n \partial^m_j\partial_{j m}+n \frac{M^2}{\hbar^2} \right)
\psi(x_1;\;\dots\; ; x_n)=0 \quad.
\ee
The mass of a single particle is given by $M$.
The index $j$ indicates which one of the $n$ particles
is affected and the index $m$ is the typical space-time index
in four flat dimensions. 
The use of the Latin indices in contrast
to the usual Greek indices will allow in the next section
allow a distinction between coordinates in
curved space-time (Greek) and coordinates in flat space-time (Latin).
The wave function (\ref{eq_wavefunction})
allows further for the construction of a conserved
current
\be
\sum_j^n \partial^m_j( i \psi^* \overleftrightarrow{\partial}_{j m}
\psi)=0 \quad.
\ee
This is still standard quantum mechanics, in order to
arrive at the dBB interpretation one rewrites the wave function
$\psi = P e^{iS/\hbar}$ by introducing a real amplitude 
$P(x_1;\;\dots\; ; x_n)$ and a real phase $S(x_1;\;\dots\; ; x_n)$.
Doing this, the complex equation (\ref{eq_KG0}) splits
up into two real equations
\begin{eqnarray} 
\label{eq_KG1}
2 M Q&\equiv&\sum_j^n(\partial^m_j S)(\partial_{jm} S)
-n M^2\quad,\\
\label{eq_KG2}
0&\equiv&
\sum_j^n\partial_{j m} \left(P^2 (\partial^m_j S) \right)\quad, 
\end{eqnarray}
where $Q(x_1;\;\dots\; ; x_n)$ is  the quantum potential, and $P(x_1;\;\dots\;
;x_n)$ is the pilot wave.
The first equation can be interpreted as a classical Hamilton-Jacobi
equation with the additional potential $Q$ and
the second equation takes the form of a conserved current.
The quantum potential in equation (\ref{eq_KG1}) is given from 
$\hbar$, the particle mass, and the pilot wave by
\be\label{eq_QuantPot}
Q=\sum_j^n\frac{\hbar^2}{2M} \frac{\partial^m_j \partial_{jm} P}{P}\quad.
\ee
This is the only way that the $\hbar$ enters
into the dBB theory.
In the dBB interpretation one postulates the existence
of particle trajectories $x_j^m(s)$ whose momentum
$p_j^m$ satisfies the relation
\be\label{eq_KG3}
p^m_j = M\frac{d x_j^m}{ds}\equiv-\partial^m_j S \quad.
\ee
Now one can derive this expression with respect to $ds$
and use the identity
\be\label{eq_der1}
\frac{d}{ds}=\sum_j^n \frac{dx^m_j}{ds}\partial_{jm} \quad.
\ee
This gives the equation of motion for all $n$ relativistic particles 
in the dBB interpretation
\be\label{eq_eom}
\frac{d^2x_j^m}{ds^2}=\sum_i^n\frac{(\partial^l_iS)(\partial_j^m\partial_{il}
S)}{M^2}\quad.
\ee
By using equation (\ref{eq_KG1}) this can be further simplified to 
\be\label{eq_eom1}
M\frac{d^2x^m_j}{ds^2}=\partial^m_j Q \quad.
\ee
The infinitesimal parameter $s$ is not
necessarily time, because every single particle
carries its own reference frame. 
For convenience one might try to choose $s$
as the eigen-time of the particle which is
finally subject to a measurement.
The non-local nature of the dBB theory
becomes obvious in the above equations of motion.
The trajectory of the particle $j$ is determined
from the potential $Q(x_1,\dots, x_n)$, which depends
on the positions of all other particles of the system.

The equations (\ref{eq_KG2}-\ref{eq_eom}) are the building 
blocks of the many particle dBB theory.
The functions $P$, $S$, and $Q$ that appear in those equations
depend on the $4\times n$ coordinates $x_j^m$.
It is therefore possible to introduce a single $4n$ dimensional
coordinate 
\be\label{eq_coordFlat}
x^L=(x_1^0,x_1^1,x_1^2,x_1^3;\;\dots\;;x_n^0,x_n^1,x_n^2,x_n^3)\quad,
\ee
which has a capital Latin index and contains the space-time positions of all $n$
particles.
One further observes that in all equations every summation
over a particle index $j$ is accompanied by a summation
over the space-time index $m$ of the corresponding particle. This allows
to replace $\partial_{jm}\rightarrow \partial_L$
and $\partial_{j}^m\rightarrow \partial^L$.
Thus, one can rewrite the equations for the
many particle case (\ref{eq_KG1}-\ref{eq_eom}) as
\begin{eqnarray}
\label{eq_KG1L}
2 M Q&\equiv&(\partial^L S)(\partial_{L} S)-n M^2\quad \mbox{with}\\
\nonumber
Q&\equiv&\frac{\hbar^2}{2M} \frac{\partial^L \partial_L P}{P}
\quad,\\
\label{eq_KG2L}
0&\equiv&
\partial_{L} \left(P^2 (\partial^L S) \right)\quad, \\ 
\label{eq_KG3L}
p^L& \equiv& M\frac{d x^L}{ds}\equiv-\partial^L S \quad, \\ 
\label{eq_eomL}
\frac{d^2x^L}{ds^2}&=&\frac{
(\partial^N S)(\partial^L\partial_N
S)}{M^2}\quad\mbox{with}\\ \nonumber
\frac{d}{ds}&\equiv& \frac{dx^L}{ds}\partial_{L} \quad.
\end{eqnarray}

\boldmath
\section{A $4\times n$ dimensional theory of curved space-time} \label{sec4n}
\unboldmath

We will now show that the equations of the
many particle dBB theory (\ref{eq_KG1L}-\ref{eq_eomL})
have a dual description in a scalar theory of
curved space-time in $4\times n$ dimensions.
As generalization of the previous single particle approach
\cite{Koch:2008rh} we will
define a setup where the momentum of every particle
is defined in the particles own four dimensional space-time.
Such a $n$-particle theory is therefore
defined in a $4\times n$ dimensional space-time.
Following the notation in (\ref{eq_coordFlat})
the coordinates in curved space-time will be denoted as
\be\label{eq_coordCurved}
\hat{x}^\Lambda=
(\hat{x}_1^0,\hat{x}_1^1,\hat{x}_1^2,\hat{x}_1^3;\;\dots\;;\hat{
x } _n^0 , \hat{x}_n^1 , \hat{x}_n^2 , \hat{x}_n^3)\quad .
\ee
The theory can be formulated by starting from the 
n-particle action
\begin{eqnarray}\nonumber
{\mathcal{S}}\left[\hat{g}_{\Lambda \Delta}\right]&=&\iint  d\alpha
\,d\hat{x}^{4n}
\sqrt{|\hat{g}|}\\
\label{eq_action}&&\;\;
 \alpha \left\{ \hat{R}+\kappa
\hat{{\mathcal{L}}}_M-{\mathcal{P}}_s
 (\hat{R}+\kappa \hat{{\mathcal{L}}}_M) \right\} \;.
\end{eqnarray}
Here, $\hat{g}$ is the determinant of the metric,
$\alpha$ is a Lagrange multiplier,
${\mathcal{P}}_s$ is a symmetrization operator
between different particles $x_i^\lambda$ and $x_j^\lambda$,
$\hat{R}$ is the Ricci scalar, $\hat{{\mathcal{L}}}_M$ is the 
matter Lagrangian, and
$\kappa$ is the coupling constant of this theory.
A variation of (\ref{eq_action}) with respect
to $\alpha$ ensures that only the
part of $\hat{R}+\kappa \hat{{\mathcal{L}}}_M$ which is symmetric under the
exchange of two particle coordinates $x_i^\lambda$
and $x_j^\lambda$ survives
\be\label{eq_sym}
{\mathcal{P}}_s\left[\hat{R}+\kappa \hat{{\mathcal{L}}}_M \right]=\hat{R}+\kappa
\hat{{\mathcal{L}}}_M\quad.
\ee
This means that
that the fields that are a solution of the problem have to be symmetric
under exchange of two particle coordinates $x_i^\lambda$
and $x_j^\lambda$. It also means
that all the different particles agree on their definition
of what is the $x_j^\mu$ direction. Therefore,
if one wants to perform a coordinate transformation
in a single four dimensional subspace, one has to perform
the same transformation in all other four dimensional subspaces.
After having derived the symmetry relation (\ref{eq_sym})
the remaining action is
\be
{\mathcal{S}}\left[\hat{g}_{\Lambda \Delta}\right]=\int 
\,d\hat{x}^{4n}
\sqrt{|\hat{g}|}
 \left( \hat{R}+\kappa
\hat{{\mathcal{L}}}_M \right) \quad.
\ee
In order to describe the local conformal part of this theory
separately one splits the metric $\hat{g}$ up into a conformal
function $\phi(x)$ and a non-conformal part $g$
\be\label{eq_gmn}
\hat{g}_{\Lambda \Gamma}=\phi^{\frac{2}{2n-1}} g_{L G}\quad.
\ee
The index notation for tensors used 
here is explained in table (\ref{tab_IC}) and
it allows to write the equations in $D=4\times n$ dimensions
either with one index (capital letters), 
or with two indices (lower case letters).
A further distinction is made between indices that
are shifted by the metric $\hat{g}$ (Greek)
and indices that are shifted by the metric $g$ (Latin).
\renewcommand\arraystretch{1.5}
\begin{table}[h]
\begin{tabular}{|c|c|c|}
\hline
Index notation & Shifted by $\hat{g}$ & Shifted by $g$ \\
\hline 
Single index & $\hat{X}_\Lambda; \dots | \{\Lambda :1 \dots
4n\} $ & $X_L; \dots |
\{L :1 \dots 4n\}$ \\
\hline
Double index & $\hat{X}_{i \lambda};\dots | \{i:1\dots \,n\,\,\}$ &
$X_{il};\dots |
\{i:1\dots \,n\,\,\}$ \\ \hline
\end{tabular}
\caption{\label{tab_IC} Index convention}
\end{table}
\renewcommand\arraystretch{1.0}
In this section we will only use the single index
notation, but all results can be immediately translated
into the double index notation used in the previous section.
The inverse of the metric (\ref{eq_gmn}) is
\be\label{eq_gmnInv}
\hat{g}^{\Lambda \Gamma}=\phi^{-\frac{2}{2n-1}} g^{L G}\quad.
\ee
Indices with a lower
Greek and a lower Roman index can be identified
$\hat{\partial}_\Lambda\equiv \partial_L$.
From this follows for example that the adjoint derivatives are not identical,
in both notations
\be\label{eq_derivatives}
\hat{\partial}^\Lambda=\hat{g}^{\Lambda \Sigma}
\hat{\partial}_\Sigma=\phi^{-\frac{2}{2n-1}}
g^{L S}\partial_S=
\phi^{-\frac{2}{2n-1}}\partial^L\quad.
\ee

\subsection{The geometrical dual to the first dBB equation}
The definition of the metric (\ref{eq_gmn})
allows to reformulate the action 
in terms of the separate functions $\phi$ and $g_{L D}$
\cite{HelayelNeto:1999tm}.
Keeping in mind the symmetrization condition (\ref{eq_sym})
the action (\ref{eq_action}) reads
\begin{eqnarray}\label{eq_action2}
{\mathcal{S}}\left[\phi,g_{L D}\right]&=&\int dx^{4n}
\sqrt{|g|} \\ \nonumber
&&
 \left[\frac{2(4n-1)}{1-2n} (\partial^L \phi)(\partial_L \phi)+ \phi^2(R+\kappa
{\mathcal{L}}_M)\right]\;.
\end{eqnarray}
We are primarily interested in studying
a flat Minkowski background space-time $g_{LD}=\eta_{LD}$.
This can be achieved by adding an additional Lagrange condition
term to the action (\ref{eq_action}) that demands a vanishing 
Weyl curvature
\begin{eqnarray}\label{eq_Nord2}
C_{\Sigma\Lambda\Xi\Delta}&=&R_{\Sigma\Lambda\Xi\Delta}-\frac{1}{2n-1}(g_{\Sigma
[
\Xi } R_ {
\beta ] \Lambda}
-g_{\Lambda[\Xi}R_{\Delta]\Sigma})\nonumber \\
&&+\frac{1}{(4n-1)(2n-1)}Rg_{\Sigma[\Xi}g_{ \Delta ] \Lambda }\\ \nonumber
&=&0\quad.
\end{eqnarray}
Such a condition also appears in the scalar theories of
curved space-time suggested by Gunnar Nordstr{\"o}m
\cite{Nordstrom:1913a,Nordstrom:1913b,Ravndal:2004ym}.
Like in standard general relativity
one further imposes that the metric has a vanishing covariant derivative
\be\label{eq_metcon}
(\nabla_\Sigma g)_{\Lambda \Xi}=0\quad.
\ee
From the conditions (\ref{eq_Nord2}, \ref{eq_metcon}) it follows
that the metric $g_{LG}$ has only $\pm 1$ on the diagonal, 
while all other entries vanish
\be
g_{LG}=\eta_{LG}\quad.
\ee
Thus, $|g|=1$, $R=0$, and therefore
the action (\ref{eq_action2}) simplifies to
\be\label{eq_action3}
{\mathcal{S}}\left[\phi\right]=\int dx^{4n}
\left[\frac{2(4n-1)}{1-2n} (\partial^L \phi)(\partial_L \phi)+\kappa \phi^2
{\mathcal{L}}_M\right]\,.
\ee
The Euler-Lagrange equation for this action is
\be\label{eq_EulLagr}
\frac{2(4n-1)}{1-2n}\frac{\partial^L\partial_L \phi}{\phi}=
\kappa{\mathcal{L}}_M\quad.
\ee
An Extension of the Hamilton Jacobi matter
Lagrangian $\hat{{\mathcal{L}}}_M$ can be defined 
by subtracting a mass term
$\hat{M}^2$ for every particle on finds
\begin{eqnarray}\label{eq_LM}
\hat{\mathcal{L}}_M &=& \hat{p}^\Lambda \hat{p}_\Lambda - 
n \hat{M}_G^2\\ \nonumber
&=&(\hat{\partial}^{\Lambda}S_H)(\hat{\partial}_{\Lambda}
S_H)-n \hat{M}_G^2\\ \nonumber
&=&\phi^{\frac{-2}{2n-1}}\left(({\partial}^{L}S_H)({\partial}_{L}
S_H)-n M_G^2\right)\\ \nonumber
&=&\phi^{\frac{-2}{2n-1}}{\mathcal{L}}_M\quad.
\end{eqnarray}
The Hamilton principle function $S_H$ defines the local momentum
$\hat{p}^\Lambda=\hat{M_G}
\,d\hat{x}^\Lambda/d\hat{s}=-\hat{\partial}^\Lambda
S_H$.
Plugging this Lagrangian into equation (\ref{eq_EulLagr}) gives
\be\label{eq_Nord1b}
\frac{2(4n-1)}{\kappa(1-2n)}\frac{\partial^L\partial_L \phi}{\phi}=
 ({\partial}^{L}S_H)({\partial}_{L}
S_H)-n M_G^2
\ee
Now one can see that this is exactly the first dBB equation (\ref{eq_KG1L})
if one identifies
\begin{eqnarray}\label{eq_match}
\phi(x)&=&P(x)\quad, \\
S_H(x)&=& S(x)\;\;, \nonumber \\
\kappa&=&\frac{2(4n-1)}{1-2n}/\hbar^2\;\;, \nonumber \\
M&=&M_G^2\quad. \nonumber
\end{eqnarray}
Note that the matching conditions demand like in the single
particle case \cite{Koch:2008rh} a negative coupling $\kappa$.

\subsection{The geometrical dual to the third dBB equation}
The stress-energy tensor corresponding to the matter
Lagrangian (\ref{eq_LM}) is
\be\label{eq_TMN}
\hat{T}^{\Lambda \Delta}=-2 (\hat{\partial}^\Lambda S_H)
(\hat{\partial}^\Delta S_H)+\hat{g}^{\Lambda \Delta} 
\left(
(\hat{\partial} S_H)^2-n\hat{M}_G^2
\right)\;.
\ee
According to the Hamilton-Jacobi formalism
the derivatives of the Hamilton principle function ($S_H$)
define the momenta 
\be\label{eq_dualKG3L}
\hat{p}_\Lambda\equiv -(\hat{\partial}_\Lambda S_H)\quad.
\ee
Therefore, with the prescription (\ref{eq_derivatives}) and
the matching condition (\ref{eq_match}) one sees immediately
that the third Bohmian equation (\ref{eq_KG3L})
is fulfilled.

\subsection{The geometrical dual to the second dBB equation}
In order to find the dual to the second Bohmian
equation one has to exploit that the stress-energy tensor (\ref{eq_TMN})
is covariantly conserved
\be
\hat{\nabla}_\Lambda \hat{T}^{\Lambda \Delta}=0\quad.
\ee
This is true if the following relations
are fulfilled
\begin{eqnarray}\label{eq_cons1}
\hat{\nabla}_\Lambda (\hat{\partial}^\Lambda S_H)&=&0 \quad,\\
\label{eq_cons2}
(\hat{\partial}^\Lambda S_H) 
\hat{\nabla}^\Delta (\hat{\partial}_\Lambda S_H)&=&0 \quad, \\
\label{eq_cons3}
(\hat{\partial}^\Lambda S_H) 
\hat{\nabla}_\Lambda (\hat{\partial}^\Delta S_H)&=&0 \quad.
\end{eqnarray}
In addition to the covariant conservation of momentum (\ref{eq_cons1})
and the conservation of squared momentum (\ref{eq_cons2}) the tensor
nature of (\ref{eq_TMN}) also demands (\ref{eq_cons3}), which
will be important in the next subsection.
In order to calculate the covariant derivatives in
(\ref{eq_cons1}-\ref{eq_cons3}),
one needs to know the 
Levi Civita connection
\begin{eqnarray}\label{eq_Levi}
\Gamma^\Sigma_{\Lambda \Delta}&=&\frac{1}{2}g^{\Sigma \Xi}\left(
\partial_\Lambda g_{\Delta \Xi}+\partial_\Delta g_{\Xi \Lambda}-
\partial_\Xi g_{\Lambda \Delta}\right) \\ \nonumber 
&=&\frac{1}{2} \phi^{-\frac{2}{2n-1}}\left[(\partial_L 
\phi^{\frac{2}{2n-1}})\delta^S_D
+(\partial_D  \phi^{\frac{2}{2n-1}})\delta^S_L\right.\\ \nonumber
&&\left.\quad\quad \;\quad-(\partial^S  \phi^{\frac{2}{2n-1}})\eta_{LD}
\right]\quad.
\end{eqnarray}
With this, the condition (\ref{eq_cons1}) reads
\be\label{eq_dualKG2L}
\hat{\nabla}_\Lambda (\hat{\partial}^\Lambda S_H)=
\phi^{-\frac{4n}{2n-1}}\partial_L \left[
\phi^2 (\partial^L S_H)
\right]=0\quad.
\ee
With the matching conditions (\ref{eq_match}), 
the above equation is identical to the second
Bohmian equation (\ref{eq_KG2L}).

\subsection{The geometrical dual to the dBB equation of motion}

The total derivative (\ref{eq_der1}) is generalized to
\be\label{eq_der2}
\frac{d}{d\hat{s}}=\frac{d\hat{x}^\Lambda}{d\hat{s}}\hat{\partial}_\Lambda=
\phi^{2/(1-2n)} \frac{dx^L}{ds}\partial_L=
\phi^{2/(1-2n)}\frac{d}{ds}\quad.
\ee
Applying this to the momentum 
 $\hat{M}_G (d\hat{x}^\Lambda/d\hat{s})=(\hat{\partial}^\Lambda S_H)$
gives the
equation of motion
\be\label{eq_eom2}
\hat{M}\frac{d^2 \hat{x}^\Lambda}{d\hat{s}^2}=
\frac{1}{\hat{M}}
(\hat{\partial}^\Delta S_H)\hat{\partial}_\Delta(\hat{\partial}^\Lambda
S_H)\quad,
\ee
or equivalently the equation of motion in the Minkowski coordinates $x^L$
\be\label{eq_eom22}
\frac{d^2x^L}{ds^2}=\frac{
(\partial^N S_H)(\partial^L\partial_N
S_H)}{M^2} \quad.
\ee
Using (\ref{eq_match}) one sees that the equation of motion (\ref{eq_eom22}) 
is dual to the equation of motion of relativistic Bohmian
mechanics (\ref{eq_eomL}). 
This is however almost a triviality because the
equations (\ref{eq_eomL}), (\ref{eq_eom2}), and (\ref{eq_eom22}) are
all derived from the same mathematical prescription (\ref{eq_der1}).
But in curved space-time there is an other equation of motion
in addition to the equation of motion (\ref{eq_eom2}), the 
geodesic equation
\be\label{eq_eom3}
\frac{d^2 \hat{x}^\Lambda}{d\hat{s}^2} +
 \hat{\Gamma}^\Lambda_{\Delta \Sigma}
\frac{d\hat{x}^\Delta}{d\hat{s}}\frac{d\hat{x}^\Sigma}{d\hat{s}}=
\hat{p}^\Lambda \cdot f(\hat{x}) \quad.
\ee
Here, $f(\hat{x})$ is some arbitrary scalar function.
For this equation of motion it is on the contrary not trivial
to show that it is consistent with (\ref{eq_eomL}).
To proof this, one can proceed a follows:
First one uses on the left hand
side of (\ref{eq_eom3}) the equations (\ref{eq_eom2}) 
and (\ref{eq_Levi}).
By using the relations (\ref{eq_cons1}) and (\ref{eq_cons3})
one can show that the result is some scalar function
times $(\hat{\partial}^\Lambda S_H)$.
This proofs that the geodesic equation  (\ref{eq_eom3}) is consistent
with equation (\ref{eq_eom2}). 
Since it was already shown that (\ref{eq_eom2}) is dual to (\ref{eq_eomL}),
one can conclude that also (\ref{eq_eom3}) is dual to (\ref{eq_eomL}).\\
Thus, all the equations
of the dBB theory (\ref{eq_KG1L}-\ref{eq_eomL})
have a dual description  
(\ref{eq_Nord1b}, \ref{eq_dualKG3L}, \ref{eq_dualKG2L}, and \ref{eq_eom3}) 
in the presented theory of
higher dimensional curved space-time.
%
\section{Interaction with an external field}
Now, the
results of the previous sections will be generalized
to interactions with an external electromagnetic field.
%
\subsection{An external field in the dBB theory}
Coupling $n$ bosonic particles (\ref{eq_KG0}) 
with charge $e$ to an external 
electromagnetic field $A_m$ is achieved by
replacing the partial derivative 
with a gauge covariant derivative
$\partial_m \rightarrow \partial_m + i e A_m/\hbar$
in the Klein-Gordon Lagrangian.
The resulting equation of motion is
\be\label{eq_KG0c}
\sum_j^n\left[ \left(
\partial^m_j\partial_{j m}+\frac{M}{\hbar^2}-
 \frac{e^2 A_j^2}{\hbar^2}\right)\psi+ 
\frac{i\partial^m_j (\psi^2 A_{jm})}{\hbar}
\right]
=0\;,
\ee
where $A_j^m$ is the electromagnetic potential $A^m$ evaluated
at the position of particle $j$.
By again rewriting the $n$-particle
wave function $\psi=P \exp (iS)$ the two
equations (\ref{eq_KG1}, \ref{eq_KG2}) generalize to
\begin{eqnarray} 
\label{eq_KG1c}
2MQ&\equiv&\sum_j^n(\partial^m_j S+eA_j^m)(\partial_{jm} S+eA_{jm})
-n M^2 ,\;\;\;\\
\label{eq_KG2c}
0&\equiv&
\sum_j^n\partial_{j m} \left(P^2 (\partial^m_j S+eA^m_j) \right)\quad, 
\end{eqnarray}
where the quantum potential $Q$ in the first equation is given by
(\ref{eq_QuantPot}).
In the presence of an external force, the
dBB definition for the particles momentum (\ref{eq_KG3})
now contains the canonical momentum $\pi_j^m$ instead
of the normal momentum $p_j^m$
\be\label{eq_KG3c}
\pi^m_j = M\frac{d x_j^m}{d\tau}\equiv-(\partial^m_j S+e A_j^m) \quad.
\ee
Thus, using (\ref{eq_der1}, \ref{eq_KG1c}, \ref{eq_KG3c},
and the relation $\partial_{kn}A_j^m=\delta_{kj} \partial_n A_j^m$)
one finds the equation of motion for all $n$ particles
in an external field $A^m$
\be\label{eq_eom1c}
M\frac{d^2x^m_j}{ds^2}=\partial^m_j Q + e
 \pi_{j n} F^{mn}\quad.
\ee
Here, the field strength tensor is
$F^{mn}_{jk}=\partial^m_j A^n_k-\partial^n_k A^m_j=\delta_{jk}F^{mn}$ 
with $F^{mn}=\partial^m A^n-\partial^n A^m$.
On the RHS of equation (\ref{eq_eom1c}) appear two terms.
The first term is the quantum potential also present in equation (\ref{eq_eom1})
and the second term is the Lorentz force which is
familiar from classical electrodynamics.

Now on can apply again the formal rewriting of coordinates (\ref{eq_coordFlat})
and the interacting equations (\ref{eq_KG1c}-\ref{eq_eom1c})
read
\begin{eqnarray}
\label{eq_KG1Lc}
2 M Q&\equiv&(\partial^L S+eA^L)(\partial_{L}S+eA_L)-n M^2\quad \\
\label{eq_KG2Lc}
0&\equiv&
\partial_{L} \left(P^2 (\partial^L S+eA^L) \right)\quad, \\ 
\label{eq_KG3Lc}
p^L& \equiv& M\frac{d x^L}{ds}\equiv-(\partial^L S+eA^L) \quad, \\ 
\label{eq_eomLc}
M\frac{d^2x^L}{ds^2}&=&\partial^L Q +e \pi_K F^{LK} \quad.
\end{eqnarray}
\subsection{An external field in the $4\times n$ dimensional theory of curved
space-time}
In the interacting case, the classical
$4 \times n$ dimensional theory of curved space-time
is analogous to the discussion in section \ref{sec4n}.
The only difference appears 
in the definition of the canonical momentum
\be\label{eq_dualKG3Lc}
\hat{\pi}^\Lambda=\hat{M}_G \frac{d \hat{x}^\Lambda}{d\hat{s}}
= - (\hat{\partial}^\Lambda S_H + e \hat{A}^\Lambda) \quad,
\ee
instead of the free momentum $\hat{p}^\Lambda$.
With this replacement the equations 
(\ref{eq_Nord1b},\ref{eq_dualKG2L},and \ref{eq_dualKG3L})
transform to
\begin{eqnarray}
\nonumber
\frac{2(4n-1)}{\kappa(1-2n)}\frac{\partial^L\partial_L\phi}{\phi}
&\equiv&(\partial^L S_H+eA^L)(\partial_{L}S_H+eA_L)\,\quad \quad \\
\label{eq_dualKG1Lc}
&& \quad \quad\quad \quad -nM^2_G\quad, \\
\label{eq_dualKG2Lc}
0&\equiv&
\phi^{\frac{4n}{1-2n}}\partial_{L} \left(P^2 (\partial^L S_H+eA^L)
\right)\,  \quad.
\end{eqnarray}
One immediately sees that with the identifications
(\ref{eq_match}) the above equations (\ref{eq_dualKG3Lc}-\ref{eq_dualKG2Lc})
are dual to the equations (\ref{eq_KG1Lc}-\ref{eq_KG3Lc}) 
In order to check the equation of motion (\ref{eq_eomLc})
we use (\ref{eq_der2}) and find
\be
\frac{d^2\hat{x}^\Lambda}{d\hat{s}^2}=
\phi^{\frac{4}{1-2n}}\frac{d^2 x^L}{ds^2}
+\hat{\pi}^L \cdot (\dots) \quad.
\ee
Using this and the identity (\ref{eq_cons3})
in its canonical version
$\label{eq_cons3c}
\hat{\pi}^\Lambda \hat{\nabla}_\Lambda \hat{\pi}^\Delta =0$
one can verify that the geodesic equation
\be\label{eq_eom3c}
\frac{d^2 \hat{x}^\Lambda}{d\hat{s}^2} +
 \hat{\Gamma}^\Lambda_{\Delta \Sigma}
\frac{\hat{\pi}^\Delta}{M_G}\frac{\hat{\pi}^\Sigma}{M_G}=
 \hat{\pi}^\Lambda \cdot (\dots)\quad,
\ee
is consistent with (\ref{eq_eomLc}).
\section{The non-relativistic limit}
The dBB theory was originally developed
for non relativistic quantum mechanics.
It will now be shown how this limit
can be obtained from the interacting relativistic
version and its dual.
Since in this limit spatial and temporal derivatives
are treated differently it is instructive
to return to the two-index notation.
Nevertheless translation to the one-index notation is
always possible with the help the index convention 
(\ref{tab_IC}). 
Starting from the relativistic and interacting many particle equations
(\ref{eq_KG1c}-\ref{eq_KG3c})
the non-relativistic limit can be obtained from
a number of low energy approximations.
At first, one wants to achieve that the
quantum phase depends only on one single
time coordinate (which will be denoted $t_1$) instead of the $n$ time
coordinates $t_j$. Therefore, 
the quantum phase S (or equivalently the
Hamilton
principal function $S_H$) is redefined by
\be\label{eq_nonr0}
S(t_j,\vec{x}_j)\equiv - M t_j+\tilde{S}(t_1,\vec{x}_j)\quad.
\ee
For the vector potential $A_j^m$ one chooses the Coulomb
gauge and imposes a similar condition for the time
coordinates and indices
\be\label{eq_nonr1}
A_j^0(t_k,\vec{x}_k)=A^0(t_1,\vec{x}_k)\,\;\mbox{and}\;A_j^i=0\quad.
\ee
For the pilot wave, the condition for the time coordinate
is analogously 
\be\label{eq_nonr2}
P(t_j,\vec{x}_j)\equiv P(t_1,\vec{x}_j)\quad.
\ee
Now the equations (\ref{eq_KG1c}, \ref{eq_KG2c}) read
\begin{eqnarray}
Q(P(t_1,\vec{x}_j))&=&-\dot{\tilde{S}}+\frac{\dot{\tilde{S}}^2}{2M}
-e A_0+\frac{e^2A_0^2}{2M}\\ \nonumber
&&-
\sum_{j=1}^n\frac{(\vec{\partial}_j\tilde{S})^2}{2M }
\; , \\
0&=&(\partial_{0}P^2)
\left( 1- \frac{eA_0}{M}-\frac{\dot{\tilde{S}}}{M}\right)\\
\nonumber
&&-P^2\frac{e\dot{A}_0+\ddot{\tilde{S}}}{M}
+\sum_{j=1}^n \vec{\partial}_{j}
\left( P^2\frac{\vec{\partial}_{j}\tilde{S}}{M}\right) \, ,
\end{eqnarray}
where derivatives with respect to $t_1$ are denoted
with a dot.
The non-relativistic limit consists now in
assuming that the mass $M$ is a much larger
quantity than all the time components of 
the four vectors
\be\label{eq_nonr3}
M\gg \dot{\tilde{S}}\,,\;M\gg eA_0\,,\; M \gg \frac{\partial_{0}P}{P}
\quad.
\ee
In this limit the above equations give
\begin{eqnarray}
0&=&
\dot{\tilde{S}}
+e A_0+
\sum_{j=1}^n\left(
\frac{(\vec{\partial}_j\tilde{S})^2}{2M }
-\hbar^2
\frac{\vec{\partial}_j^2P}{2P M}
\right)
\; , \\
0&=& \partial_{0}(P^2)
+ \sum_{j=1}^n\vec{\partial}_{j}
\left( P^2\frac{\vec{\partial}_{j}\tilde{S}}{M}\right) 
\quad.
\end{eqnarray}
By replacing $\psi=P e^{i\tilde{S}/\hbar}$
one sees that this is the many particle Schr{\"o}dinger
equation with the potential $eA_0$.
Finally, with the same relations (\ref{eq_nonr1},\ref{eq_nonr3})
the equation (\ref{eq_KG3c}) gives the original
momentum definition in the Bohmian interpretation
\be
\vec{p}_j=\vec{\partial}_j\tilde{S} \quad.
\ee

Due to the matching conditions (\ref{eq_match}),
in the geometrical dual the
non-relativistic limit
means the following:
First, the Hamilton principle function $S_H$ is split
up into a linear part containing the mass and a remaining
part $\tilde{S}_H$ (\ref{eq_nonr0}).
Second, all the time coordinates $t_j$ in
the functions $\tilde{S}_H$, $\phi$, and $A$ are 
synchronized (\ref{eq_nonr0}-\ref{eq_nonr2}) and
the Coulomb gauge is chosen.
Third, the time components of the remaining
four vectors are taken to the limit where
they are much smaller than the mass $M_G$ (\ref{eq_nonr3}).\\
In this limit one finds in the geometrical theory
that the function $\phi$ plays a double role.
On the one hand it is a conformal field of
the metric which interacts with matter.
On the other hand its square $\phi^2(t,\vec{x}_j)$ 
gives the probability
density of finding the particles of the system
in the positions $\vec{x}_j$ at the time $t$.
When one intends to find normalizable solutions
(for instance of a central potential $A_0\sim 1/r$)
usually one imposes
the boundary condition $lim_{\vec{x}\rightarrow \infty}\phi^2=0$.
This limit means that in the dual theory the
metric $\hat{g}_{\Lambda \Delta}$ as to vanish
asymptotically for far regions.

\section{Discussions}
In this section some interpretational 
issues are addressed. This aims to develop a physical
understanding of the presented duality.
\subsection{Locality}
Quantum mechanics in the dBB interpretation
is a theory which allows to talk about particle
position at the cost of non-local interactions.
The non-locality becomes obvious 
by looking at the equation of motion
for a free particle in the dBB theory (\ref{eq_eom1}).
The movement of this ``free'' particle is not free
but it is governed by
the quantum potential $Q$, which simultaneously
depends on the positions of all the other
particles of the system. It is often
argued that such a non-locality can not
have any dual in a (by construction) local theory such
as a geometrical theory of curved space-time.
Therefore the first question is:

``How can a non-local theory have a dual local theory?''\\
The reason is that in the presented theory of curved
space-time, every single particle is living in
its own four dimensional space-time. Therefore,
the positions of $n$ different particles correspond
to one single point in the $4\times n$ dimensional
space-time. 
In some sense the introduction of additional dimensions
seems to be a common feature of a classical
understanding of quantum theories (see the AdS/CFT dualities or
the many worlds interpretation). 
This higher dimensional construction helps around
the non-locality argument but it brings a new
problem:

\subsection{Four dimensional space-time}
``How does the $4\times n$ dimensional toy model
cope with the fact that every accepted
physical theory is formulated in four dimensions?''\\
We don't know the answer to this question yet.
Therefore, it is up to now only prooven that
the presented duality exists mathematically, but
a better physical understanding is certainly desirable.
Part of the answer might lie in the
 symmetrization condition $\mathcal{P}_s$
in the action (\ref{eq_action}), which corresponds to the
symmetrization in bosonic many particle quantum mechanics. 
Since all solutions have to be
invariant under the exchange of two sub-coordinates 
$x_{i}^\mu$ and $x_j^\mu$ it is clear
that all those sub-coordinates $x_i^\mu$ have to have
the same interpretation. This symmetry forbids for instance 
to rotate the $x$ and $y$ coordinates for a single particle 
without rotating the coordinates for the other particles.
In this sense the interpretation of common sense coordinates
in $4$ dimensions is unique.

\subsection{Gravity}
``How is this geometrical theory related to THE geometrical theory
- general relativity?''
It is very tempting to think about such a relation
because the action (\ref{eq_action}) is very similar
to the Einstein Hilbert action and the metric condition (\ref{eq_metcon})
is the same as in general relativity.
However the presented
theory in this higher dimensional form can NOT describe 
four dimensional gravity. The
possible connection between both
geometrical theories still has to be explored.

\subsection{The matching conditions}
The matching conditions (\ref{eq_match})
are not unique, but they were chosen
in order to have the most direct connection
between three functions.
Those are the quantum phase $S$ connected to the Hamilton
principle function $S_H$, the quantum amplitude $P$
connected to the conformal metric function $\phi$, 
and the quantum mass $M$ connected to the
mass $M_G$. Fixing those relations forces a
relation between the particle number $n$, the Planck
constant $\hbar$ and the coupling of the 
geometrical theory $\kappa$
\be\nonumber
\kappa=\frac{2(4n-1)}{1-2n}/\hbar^2\;\;.
\ee
This has the consequence
that if one gives $\hbar$ a fixed numerical
value, then the coupling of the geometrical
theory $\kappa$ runs from $-6/\hbar^2$ to $-4/\hbar^2$, depending
on the number of particles ($n=1, \dots ,\infty$) which are
part of the system.
Reversely demanding a fixed geometrical coupling
would result in a running Planck constant $\hbar$.
This behavior seems to be the only unexpected (and undesired)
peculiarity of the chosen matching conditions (\ref{eq_match}). 
More optimistically, one might also see the scale dependence
of the coupling $\kappa$ as a feature of the
toy model which it shares with effective quantum
field theories and even some approaches to gravity \cite{Reuter:2001ag}.

\section{Summary}

In this paper we showed
that the equations of the free relativistic dBB theory
for many particles (\ref{eq_KG1L}-\ref{eq_eomL})
have a dual description in a $4\times n$ dimensional
theory of curved space-time.
For the translation between the two theories a single
set of matching conditions was defined (\ref{eq_match}).
The result was then generalized to interactions with an
external electromagnetic field.
Before discussing interpretational issues, the
limit of the
non-relativistic Schr{\"o}dinger equation 
is derived.
The important question whether such dualities can
also be found for, self-interacting theories,
fermions, or quantum field theory will be subject of future
studies.\\ \\
The author wants to thank 
M. A. Diaz,
J.M. Isidro,
H. Nikolic, 
D. Rodriguez, 
and C. Valenzuela, 
for very helpful comments and discussions.

\begin{thebibliography}{15}
\expandafter\ifx\csname natexlab\endcsname\relax\def\natexlab#1{#1}\fi
\expandafter\ifx\csname bibnamefont\endcsname\relax
  \def\bibnamefont#1{#1}\fi
\expandafter\ifx\csname bibfnamefont\endcsname\relax
  \def\bibfnamefont#1{#1}\fi
\expandafter\ifx\csname citenamefont\endcsname\relax
  \def\citenamefont#1{#1}\fi
\expandafter\ifx\csname url\endcsname\relax
  \def\url#1{\texttt{#1}}\fi
\expandafter\ifx\csname urlprefix\endcsname\relax\def\urlprefix{URL }\fi
\providecommand{\bibinfo}[2]{#2}
\providecommand{\eprint}[2][]{\url{#2}}

\bibitem[{\citenamefont{Bohm}(1951{\natexlab{a}})}]{Bohm:1951}
\bibinfo{author}{\bibfnamefont{D.}~\bibnamefont{Bohm}}, \bibinfo{journal}{Phys.
  Rev.} \textbf{\bibinfo{volume}{85}}, \bibinfo{pages}{166}
  (\bibinfo{year}{1951}{\natexlab{a}}).

\bibitem[{\citenamefont{Bohm}(1951{\natexlab{b}})}]{Bohm:1951b}
\bibinfo{author}{\bibfnamefont{D.}~\bibnamefont{Bohm}}, \bibinfo{journal}{Phys.
  Rev.} \textbf{\bibinfo{volume}{85}}, \bibinfo{pages}{180}
  (\bibinfo{year}{1951}{\natexlab{b}}).


\bibitem{Holland:1985ud}
  P.~R.~Holland and J.~P.~Vigier,
  Nuovo Cim.\  B {\bf 88}, 20 (1985);
  P.~R.~Holland,
  Phys.\ Lett.\  A {\bf 128}, 9 (1988);
  P.~R.~Holland,
  Phys.\ Rept.\  {\bf 224}, 95 (1993).


\bibitem{Nikolic:2002mi}
  H.~Nikolic,
  Found.\ Phys.\ Lett.\  {\bf 17}, 363 (2004)
  [arXiv:quant-ph/0208185];
  H.~Nikolic,
  Found.\ Phys.\ Lett.\  {\bf 18}, 123 (2005)
  [arXiv:quant-ph/0302152];
  H.~Nikolic,
  Found.\ Phys.\ Lett.\  {\bf 18}, 549 (2005)
  [arXiv:quant-ph/0406173].



\bibitem{Struyve:2006cj}
  W.~Struyve and H.~Westman,
  AIP Conf.\ Proc.\  {\bf 844}, 321 (2006)
  [arXiv:quant-ph/0602229].

\bibitem{Nikolic:2006az}
  H.~Nikolic,
  Found.\ Phys.\  {\bf 37}, 1563 (2007)
  [arXiv:quant-ph/0609163].

\bibitem[{\citenamefont{Bell}(1988)}]{Bell:1988}
\bibinfo{author}{\bibfnamefont{J.~S.} \bibnamefont{Bell}},
  \emph{\bibinfo{title}{Speakable and Unspeakable in Quantum Mechanics}}
  (\bibinfo{publisher}{Cambridge University Press; ISBN 0521368693},
  \bibinfo{year}{1988}).

\bibitem[{\citenamefont{Kochen and Specker}(1976)}]{Kochen:1976}
\bibinfo{author}{\bibfnamefont{S.} \bibnamefont{Kochen}} \bibnamefont{and}
\bibinfo{author}{\bibfnamefont{E.~P.} \bibnamefont{Specker}},
 \bibinfo{journal}{Journal of Mathematics and Mechanics}
\textbf{\bibinfo{volume}{17}},
  \bibinfo{pages}{59} (\bibinfo{year}{1976}).

\bibitem{Koch:2008hn}
  B.~Koch,
  arXiv:0801.4635 [quant-ph].

\bibitem{Koch:2008rh}
  B.~Koch,
  arXiv:0810.2786 [hep-th].

\bibitem{Santamato:1984qe}
  E.~Santamato,
  J.\ Math.\ Phys.\  {\bf 25}, 2477 (1984);
  E.~Santamato,
  Phys.\ Rev.\  D {\bf 32}, 2615 (1985).

\bibitem{Shojai:2000us}
  F.~Shojai and A.~Shojai,
  Int.\ J.\ Mod.\ Phys.\  A {\bf 15}, 1859 (2000)
  [arXiv:gr-qc/0010012];
  A.~Shojai,
  Int.\ J.\ Mod.\ Phys.\  A {\bf 15}, 1757 (2000)
  [arXiv:gr-qc/0010013];
  F.~Shojai and A.~Shojai,
  Pramana {\bf 58}, 13 (2002)
  [arXiv:gr-qc/0109052];
  F.~Shojai and A.~Shojai,
  arXiv:gr-qc/0404102.

\bibitem{Carroll:2004hs}
  R.~Carroll,
  arXiv:gr-qc/0406004;
  R.~Carroll,
  arXiv:quant-ph/0406203;
  R.~Carroll,
  arXiv:math-ph/0701007;
  R.~Carroll,
  arXiv:0705.3921 [gr-qc].

\bibitem{Carroll:2007zh}
  R.~Carroll,
  arXiv:math-ph/0703065;
  M.~Novello, J.~M.~Salim and F.~T.~Falciano,
  arXiv:0901.3741 [gr-qc].


\bibitem{Abraham:2008yr}
  J.~M.~Isidro, J.~L.~G.~Santander and P.~F.~de Cordoba,
  arXiv:0808.2351 [hep-th];
  S.~Abraham, P.~F.~de Cordoba, J.~M.~Isidro and J.~L.~G.~Santander,
  arXiv:0810.2236 [hep-th].
  S.~Abraham, P.~F.~de Cordoba, J.~M.~Isidro and J.~L.~G.~Santander,
  arXiv:0810.2356 [hep-th].


\bibitem{HelayelNeto:1999tm}
  J.~A.~Helayel-Neto, A.~Penna-Firme and I.~L.~Shapiro,
  Phys.\ Lett.\  B {\bf 479}, 411 (2000)
  [arXiv:gr-qc/9907081].

\bibitem[{\citenamefont{Nordstrom}(1913{\natexlab{a}})}]{Nordstrom:1913a}
\bibinfo{author}{\bibfnamefont{G.}~\bibnamefont{Nordstrom}},
  \bibinfo{journal}{Ann. d. Phys.} \textbf{\bibinfo{volume}{40}},
  \bibinfo{pages}{872} (\bibinfo{year}{1913}{\natexlab{a}}).

\bibitem[{\citenamefont{Nordstrom}(1913{\natexlab{b}})}]{Nordstrom:1913b}
\bibinfo{author}{\bibfnamefont{G.}~\bibnamefont{Nordstrom}},
  \bibinfo{journal}{Ann. d. Phys.} \textbf{\bibinfo{volume}{42}},
  \bibinfo{pages}{533} (\bibinfo{year}{1913}{\natexlab{b}}).

\bibitem{Ravndal:2004ym}
  F.~Ravndal,
  arXiv:gr-qc/0405030.

\bibitem{Reuter:2001ag}
  M.~Reuter and F.~Saueressig,
  Phys.\ Rev.\  D {\bf 65}, 065016 (2002)
  [arXiv:hep-th/0110054].


\end{thebibliography}


\end{document}